\documentclass[preprint2]{aastex}
\newcommand{\Msun}{\ifmmode\mbox{M}_{\odot}\else$\mbox{M}_{\odot}$\fi}
\newcommand{\Rsun}{\ifmmode\mbox{R}_{\odot}\else$\mbox{R}_{\odot}$\fi}
\newcommand{\Mearth}{\ifmmode\mbox{M}_{\oplus}\else$\mbox{M}_{\oplus}$\fi}
\newcommand{\Rearth}{\ifmmode\mbox{R}_{\oplus}\else$\mbox{R}_{\oplus}$\fi}

\shorttitle{10 Years of {\em{RXTE}} Monitoring of AXP 4U~0142+61}
\shortauthors{Dib et al.}
\begin{document}
%% --------------------------------------------------------
\title{10 Years of {\em{RXTE}} Monitoring of Anomalous X-ray Pulsar
4U~0142+61: Long-Term Variability}
%% --------------------------------------------------------
\author{Rim~Dib\altaffilmark{1},
        Victoria~M.~Kaspi\altaffilmark{1}, and
	Fotis~P.~Gavriil\altaffilmark{2}\altaffilmark{3}}
%% --------------------------------------------------------
%% \email{}
%% --------------------------------------------------------
\altaffiltext{1}{Department of Physics, McGill University,
                 Montreal, QC H3A~2T8.}
\altaffiltext{2}{NASA Goddard Space Flight Center, Greenbelt, MD.}
\altaffiltext{3}{pak Ridge Associated Universities, Oak Ridge, TN.}
%% --------------------------------------------------------
\begin{abstract}
%% --------------------------------------------------------
We report on 10 years of monitoring of the 8.7-s Anomalous X-ray Pulsar
4U~0142+61 using the {\emph{Rossi X-Ray Timing Explorer (RXTE)}}. This
pulsar exhibited stable rotation from 2000 March until 2006 February: the
RMS phase residual for a spin-down model which includes $\nu$, $\dot{\nu}$,
and $\ddot{\nu}$ is 2.3\%. We report a possible phase-coherent timing
solution valid over a 10-yr span extending back to March~1996. A glitch may
have occured between 1998 and 2000, but is not required by the existing
timing data. The pulse profile has been evolving since 2000. In particular,
the dip of emission between its two peaks got shallower between 2002 and
2006, as if the profile were evolving back to its pre-2000 morphology,
following an earlier event, which possibly also included the glitch
suggested by the timing data. These profile variations are seen in the
2$-$4~keV band but not in 6$-$8~keV. We also detect a slow increase in the
pulsed flux between 2002 May and 2004 December, such that it has risen by
36$\pm$3\% over 2.6 years in the 2$-$10~keV band. The pulsed flux
variability and the narrow-band pulse profile changes present interesting
challenges to aspects of the magnetar model.
%% --------------------------------------------------------
\end{abstract}
%% --------------------------------------------------------
%% --------------------------------------------------------
%% A small citation guide:
%% \citep{jon90} (Jones et al. 1990) [no comma] 
%% \cite(t){jon90} Jones et al. (1990) [no comma]
%%
%% \citep*{jon90} (Jones, Baker, and Williams 1990) [one arg, no comma]
%% \citet*{jon90} Jones, Baker, and Williams (1990) [one arg]
%%
%% \citealp(jon90) Jones et al. 1990  [no comma]
%% \citealt{jon90} Jones et al. 1990  [no comma, been using this]
%%
%% \citealp*{jon90} Jones, Baker, and Williams 1990 [all and no comma,this]
%% \citealt*{jon90} Jones, Baker, and Williams 1990 [all and no comma]
%% --------------------------------------------------------
%%               __   __
%%              __ \ / __
%%             /  \ | /  \
%%                 \|/
%%            _,.---v---._
%%   /\__/\  /            \
%%   \_  _/ /              \ 
%%     \ \_|           @ __|
%%      \                \_
%%       \     ,__/       /
%%     ~~~`~~~~~~~~~~~~~~/~~~~
%% ~~~~~~~~~~~~~~~~~~~~~~~~~~~~~~~~~~  hjw97
%% --------------------------------------------------------
\section{Introduction}
%% --------------------------------------------------------
%% PARAGRAPH MAGNETARS.
The existence of magnetars -- young, isolated neutron stars powered by the
decay of an ultrahigh magnetic field -- is now well supported by many
independent lines of evidence \citep{wtastro}. This comes from the study of
soft-gamma repeaters (SGRs) and anomalous X-ray pulsars (AXPs), both of
which classically exhibit X-ray pulsations having luminosity in the range
10$^{34-36}$~erg~s$^{-1}$, periods $P$ ranging from 5-12~s, $\dot{P}$s of
10$^{-13}$$-$10$^{-11}$, and surface dipolar magnetic fields $B$ in the
range 0.6$-$7 $\times$ 10$^{14}$~G, assuming vacuum dipole magnetic
braking\footnote{Magnetic fields discussed in this paper are calculated via
$B \equiv 3.2 \times 10^{19} \sqrt{P \dot{P}}$~G, where $P$ is the spin
period in seconds and $\dot{P}$ is the period derivative.}. AXPs and SGRs,
in the magnetar model, are ultimately powered by the internally decaying
magnetic field. In the magnetar model, the pulsed X-rays are suggested to be
the result of a combination of surface thermal emission and a non-thermal
high-energy component from resonant scattering of thermal photons off
magnetospheric currents \citep*{tlk02rim}. Magnetar bursting, the hallmark
of SGRs and also seen in AXPs, is believed to be a result of crustal yield
and subsequent magnetospheric disturbances ultimately caused by stresses on
the crust by the decaying internal field (see for example
\citealt{2259outburstw}).
%% ==
\par
%% ==
%% VICKY PARAGRAPH INTRODUCING PHENOMENAE.
Recently, thanks in large part to long-term monitoring campaigns, it has
become clear that AXPs exhibit a variety of types of aperiodic X-ray
variability that can in principle be useful for testing aspects of the
magnetar model. This variability can be categorized into four types, some of
which are seen contemporaneously with each other: very short-duration
SGR-like bursts, sudden outbursts and transient brightenings with decays
lasting months or longer, slow-rise long-term flux variations also with slow
decays, and pulse profile changes.
%% ==
\par
%% ==
%% PARAGRAPH PHENOMENA 2259.
Classic examples of SGR-like bursts and an outburst were seen in 2002 for
AXP 1E~2259+586, which exhibited a sudden order-of-magnitude increase in the
pulsed and total flux, followed by a one-year-long flux decay
(\citealt{2259outburstk}; \citealt{2259outburstw}). The outburst was
accompanied by over 80 short SGR-like bursts, a rotational glitch with
interesting recovery on a time-scale of 2 weeks, short-lived spectral
changes, and dramatic broad-band pulse morphology changes which included the
two profile peaks swapping heights and which lasted 2--3 weeks. The event
was consistent with the picture of sudden crustal yield influencing both the
interior and the exterior of the AXP, in analogy with large SGR bursts.
%% ==
\par
%% ==
%% PARAGRAPH MORE PHENOMENA 2259 AND OTHER AXPS.
Several observations of this same source and others suggest similar
outbursts in AXPs that went undetected.  {\em{GINGA}} \enspace observations
\enspace of \enspace 1E~2259+586 reported by \citet*{2259ginga} also showed
a factor-of-two pulsed flux change and pulse profile variations, both of
which, in hindsight, could be explained by a contemporaneous but short-lived
outburst that went unseen.  In AXP~1RXS~J170849.0$-$4000910, two rotational
glitches were discovered (\citealp*{1708glitch1}; \citealt{1708glitch2};
\citealt{1708profilechange}).  \citet{1708profilechange} reported possible
small pulse morphology changes associated with these glitches. Whether these
glitches were accompanied by bursting that went unobserved is unknown but
plausible.  The transient AXP~XTE~J1810$-$197 underwent a dramatic sudden
brightening by nearly two orders of magnitude \citep{1810brightening}
followed by a total flux decay that lasted years \citep{1810fadingmodels};
this may well have been an outburst similar to, though larger than that in
1E~2259+586, but for which the brief main event went observed. Similarly,
the transient candidate AXP~AX~J1845$-$0258 underwent a factor $>100$ decay
in flux after an initial brightening that lead to its discovery
\citep{1845decrease, 1845cindy}.  This too could have been the result of an
unseen outburst.
%% ==
\par
%% ==
%% PARAGRAPH 1048 FLARES.
AXP outbursts appear to be fundamentally different from the slow-rise,
long-term flux variations seen in AXP~1E~1048.1$-$5937.  \citet{1048flares}
discovered two long-lived, slow-rise X-ray pulsed flux flares from this
source.  The first flare had peak pulsed flux a factor of $\sim$2 greater
than the quiescent level, and lasted $\sim$100 days.  The second, larger
flare had peak a factor of $>4$ higher than in quiescence, and lasted over
one year.  The flares were accompanied by an increase in the phase-averaged
flux of the source and a decrease in pulsed fraction, although the time
scale and full dynamic range of the these changes have not been clearly
established \citep{1048unpulsedflares, tiengo2005}. 
%%%%%% {ALSO MENTION HERE: NO FLUX-HARDNESS CORRELATION EITHER, BUT AGAIN, 
%%%%%% SAMPLED MODEST LUMINOSITY VARIATION}.
No simultaneous pulse morphology changes were detected, and though the
source did exhibit some SGR-like bursts \citep*{burststudy}, they were not
obviously correlated with pulsed flux. Large (factor of 10) torque changes
were seen especially during the large flare, but the correlation between
torque and pulsed flux, at least on time scales smaller than the flare
itself, was marginal. Overall, the slow-rise flares seen in 1E~1048.1$-$5937
are not thought to result from crustal cracking as in outbursts. They can,
however, be explained by a spontaneous increase in the magnetic field twist
in the magnetosphere.  However what might trigger such events is unclear.
Nevertheless, \citet{1708hardness} and \citet{1708campana} found, using
observations of 1RXS~J170849.0$-$400910 which has also shown phase-averaged
flux variability, that one important prediction of the twisted magnetosphere
model appears to hold, namely a correlation between spectral hardness and
flux. 
%% ==
\par
%% ==
%% PARAGRAPH 0142.
4U~0142+61 \enspace is \enspace an \enspace 8.7-s \enspace AXP. \quad It
\enspace has \enspace $\dot{P}$~=~0.2~$\times~10^{-11}$, implying a surface
dipole magnetic field of 1.3~$\times~10^{14}$~G. From continuous {\em{RXTE}}
monitoring, \citet{0142previous} showed that 4U~0142+61 rotates with high
stability. However, \citet*{0142asca} reported a possible timing glitch in
1999 on the basis of an {\em{ASCA}} observation in which the value of the
frequency was marginally discrepant with the frequency as reported by
\citet{0142previous}. \citeauthor{0142asca} also reported simultaneous pulse
morphology changes. As of 2006~April, in the published flux history of this
source, there had been no reports of any X-ray activity like the flares of
1E~1048$-$5937 and the outburst of 1E~2259+586. However, very recently, in
2006 April and June and again in 2007 February
\citep{atel1,atel2,0142lastburst} SGR-like bursts were seen from 4U~0142+61,
along with a sudden pulse profile change and a timing anomaly.  In this
paper we refer to the history prior to April 2006.
%% ==
\par
%% == 
%% PARAGRAPH REPORTING.
Here we report on continued {\em{RXTE}} monitoring observations in which we
find a possibly new type of AXP variability, namely a slow, long-term
increase in the pulsed flux accompanied by slow pulse profile evolution.  We
also report on 10 years of timing and investigate the possibility of a
glitch having occured in 1998 or 1999, during an observing gap, which may
have precipitated the evolution we are witnessing today. Our observations
are described in Section~\ref{sec:observations}. Our timing, pulsed
morphology, and pulse flux analysis are presented, respectively, in
Sections~\ref{sec:timing}, \ref{sec:profile}, and~\ref{sec:flux}. We present
a combined pulse shape and pulsed flux analysis in Section~\ref{sec:combo}.
In Section~\ref{sec:discussion}, we discuss the possible origins of this
behavior and the implications for the magnetar model. 
%% --------------------------------------------------------
\section{Observations}
\label{sec:observations} 
\begin{figure}
\includegraphics[scale=.65]{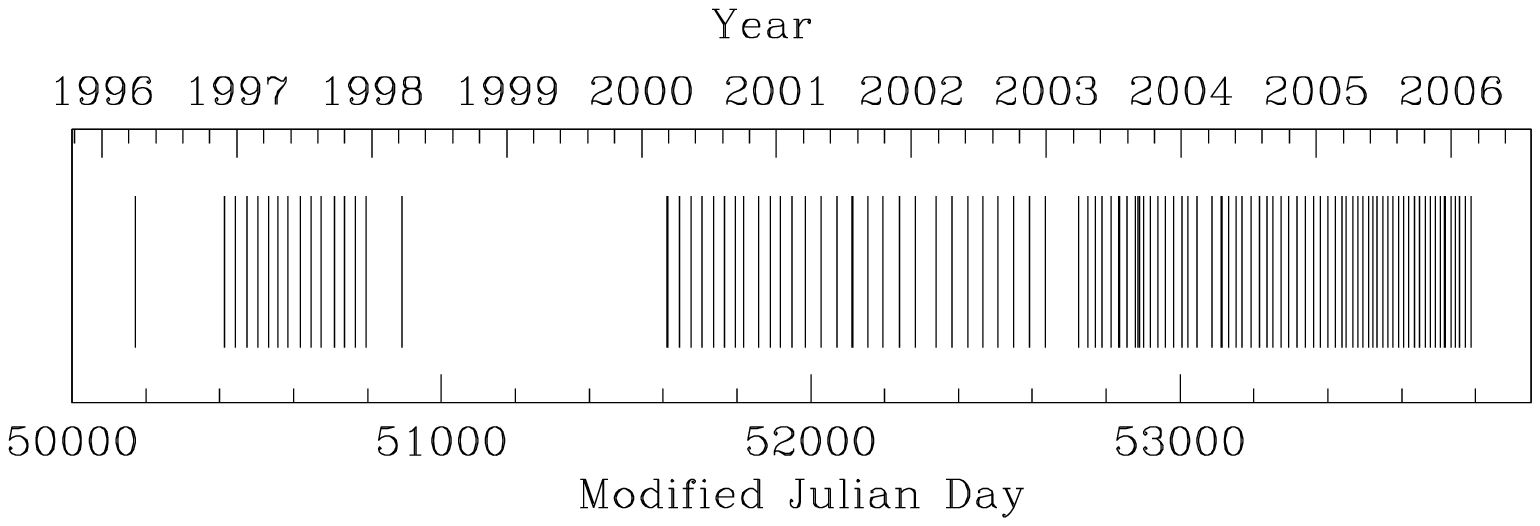}
\caption
{
Epochs of observations of 4U~0142+61 with {\em{RXTE}}.
\label{figure1}
}
\end{figure}

%% --------------------------------------------------------
%% PARAGRAPH OBSERVATIONS.
The results presented here were obtained using the Proportional Counter
Array (PCA) on board {\em{RXTE}}. The PCA consists of an array of five
collimated xenon/methane multi-anode proportional counter units (PCUs)
operating in the 2$-$60 keV range, with a total effective area of
approximately 6500~cm$^2$ and a field of view of $\sim$1$^{\circ}$ FWHM
\citep{jahoda}. Our 136 observations are of various lengths (see
Table~\ref{table1}). Most were obtained over a period of several years as
part of a long-term monitoring program, but some are isolated observations
(see Fig.~\ref{figure1}).
%% ==
\par
%% ==
%% PARAGRAPH GAP.
Note that there is a 2-year gap in the observations on which we are
reporting: no {\em{RXTE}} observations were made from 03/21/1998
(MJD~50893.083) to 03/07/2000 (MJD~51610.617). The gap exists because
4U~0142+61 was only added to our regular AXP monitoring program at the start
of {\em{RXTE}} Cycle~5. Prior to the gap, our observations consist of
a)~4~very closely spaced {\em{RXTE}}~Cycle~1 observations,
b)~14~short~Cycle~2 observations spanning a period of a year, and c)~a
single Cycle~3 observation. No observations were made in Cycle~4
(see~Table~\ref{table1}, Fig.~\ref{figure1}).
%% ==
\par
%% ==
%% PARAGRAPH DATAMODES.
For the monitoring, we used the GoodXenonwithPropane data mode except during
Cycle 10 when we used the GoodXenon mode.  Both data modes record photon
arrival times with 1-$\mu$s resolution and bin energies into one of 256
channels. To maximize the signal-to-noise ratio, we analysed only those
events from the top xenon layer of each PCU. 
%% ======
%% |    |
%% |    |  FIGURE OBSERVATIONS. [h]
%% |    |
%% ======
%% ======
%% |    |
%% |    |  TABLE OBSERVATIONS.
%% |    |
%% ======
%% --------------------------------------------------------
\section{Analysis and Results} 
\label{sec:analysis}
%% --------------------------------------------------------
\subsection{Phase-coherent Timing}
\label{sec:timing}
%% --------------------------------------------------------
%% PARAGRAPH TIMING BEGINNING. USING DS TOOLS.
Photon arrival times at each epoch were adjusted to the solar system
barycenter using the position obtained by \citet{0142position} from
{\em{Chandra}} data. They were then binned with 31.25-ms time resolution. In
the timing analysis presented below, we included only the events in the
energy range 2$-$10~keV (unless otherwise specified) to maximize the
signal-to-noise ratio of the pulse.
%% ==
\par
%% ==
%% PARAGRAPH EXPLAIN TIMING.
Each barycentric binned time series was epoch-folded using an ephemeris
determined iteratively by maintaining phase coherence; see below. 
%%%%%% {AFTER THE ITERATIONS WERE OVER, AN EPHEMERIS THAT IS THE SAME 
%%%%%% WITHIN ERRORS AS THE POST-GAP EPHEMERIS WAS USED EVERYWHERE. 
%%%%%% FIND A WAY TO SAY THAT HERE}.
Resulting pulse profiles, with 64 phase bins, were cross-correlated in the
Fourier domain with a high signal-to-noise template created by adding
phase-aligned profiles from all observations. The cross-correlation returned
an average pulse time of arrival (TOA) for each observation corresponding to
a fixed pulse phase. The pulse phase $\phi$ at any time $t$ can be expressed
as a Taylor expansion,
%%%%%%
\begin{eqnarray}
\phi(t) & = &
\phi_{0}(t_{0})+\nu_{0}(t-t_{0})+
\frac{1}{2}\dot{\nu_{0}}(t-t_{0})^{2} \nonumber \\
& & +\frac{1}{6}\ddot{\nu_{0}}(t-t_{0})^{3}+{\ldots},
\end{eqnarray}
%%%%%%
where $\nu$~$\equiv$~1/$P$ is the pulse frequency,
$\dot{\nu}$~$\equiv$~$d\nu$/$dt$, etc$.$, and subscript ``0'' denotes a
parameter evaluated at the reference epoch $t=t_0$. The TOAs were fitted to
the above polynomial using the pulsar timing software package
TEMPO\footnote{See http://www.atnf.csiro.au/research/pulsar/tempo.}.
%% ==
\par
\begin{figure}
\includegraphics[scale=.55]{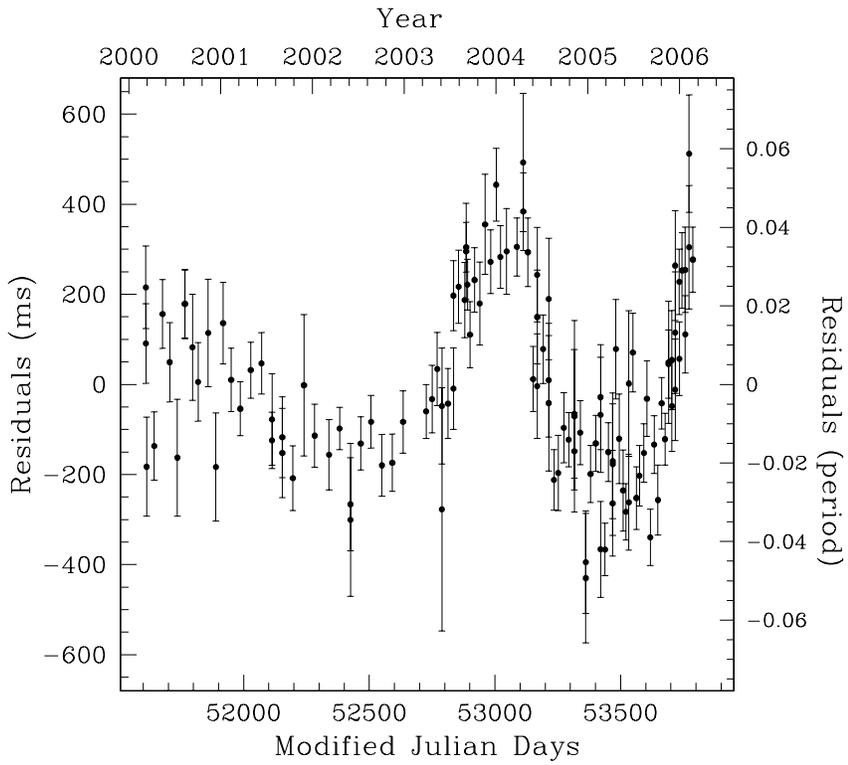}
\caption 
{ 
Arrival time residuals for 4U~0142+61 for the post-gap period, using the
post-gap ephemeris given in Table~\ref{table2}. 
The residuals have RMS 2.3\% of the pulse period. 
\label{figure2} 
}
\end{figure}

%% ==
%% PARAGRAPH POSTGAP EPHEM.
We report an unambiguous phase-coherent timing solution that spans the 
post-gap (i.e. after March 2000, MJD 51610) 6-yr period up until February
2006 (MJD 53787) including all data in {\emph{RXTE}} Cycles~5$-$10.  The
parameters of our best-fit spin-down model which includes $\nu$,
$\dot{\nu}$, and $\ddot{\nu}$ are presented in Table~{\ref{table2}}.  The
corresponding phase residuals are shown in Figure~{\ref{figure2}}. Note the
unmodelled features in the residuals; these may be caused by a noise process
similar to that commonly seen in radio pulsar timing (e.g. \citealt{1509maggie}).
%% ======
%% |    |
%% |    |  FIGURE REZPOST.[h]
%% |    |
%% ======
%% ======
%% |    |
%% |    |  TABLE EPHEMERIDES.
%% |    |
%% ======
%% ==
\par
\begin{figure}
\includegraphics[scale=.55]{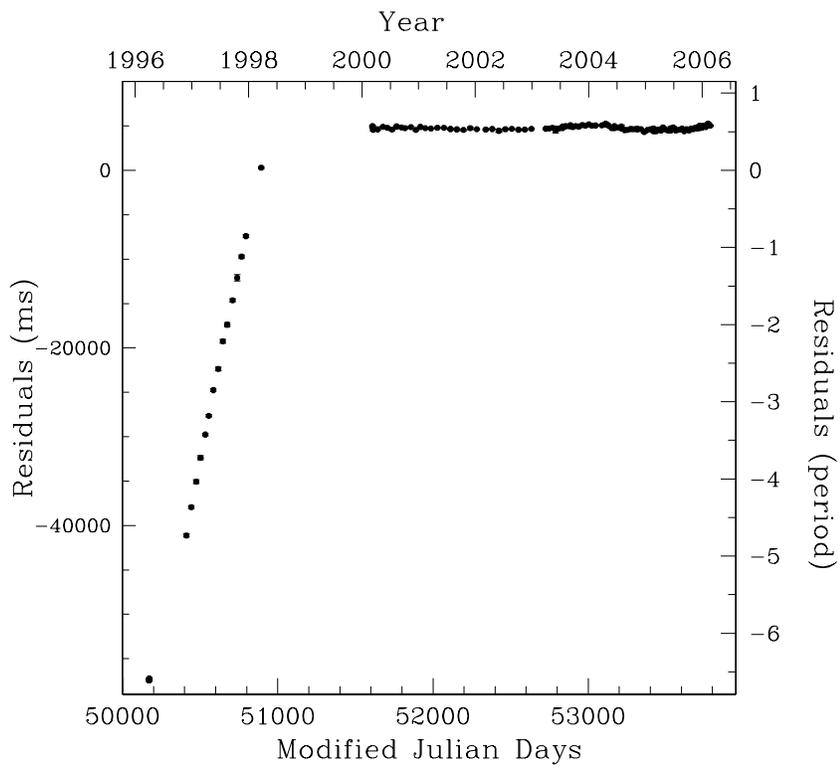}
\caption
{
Arrival time residuals for 4U~0142+61 for all {\em{RXTE}} Cycles using the
post-gap ephemeris.
\label{figure3}
}
\end{figure}
\begin{figure}
\includegraphics[scale=.55]{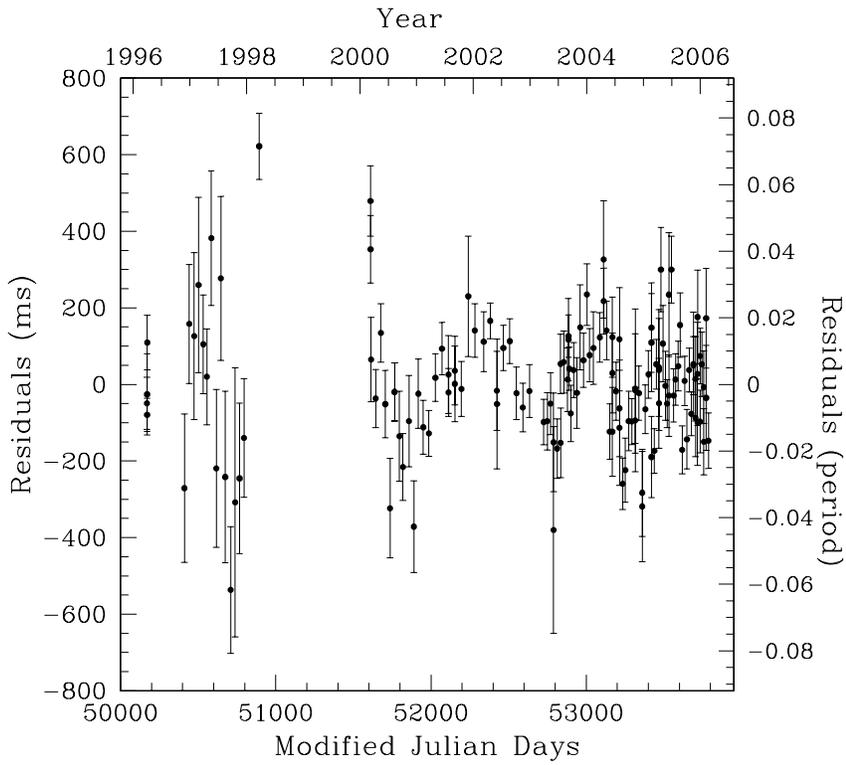}
\caption
{
Arrival time residuals for 4U~0142+61 for all {\em{RXTE}} Cycles using the
overall ephemeris (see Table~\ref{table2}). 
The residuals have RMS 1.9\% of the pulse period.
\label{figure4}
}
\end{figure}

%% ==
%% PARAGRAPH MAYBE GLITCH.
The best-fit post-gap ephemeris does not, however, fit the pre-gap TOAs
well. Figure~{\ref{figure3}} shows a clear systematic deviation in the
pre-gap residuals obtained after subtracting the post-gap ephemeris. The
best-fit frequency obtained from the post-gap model at the reference epoch
is larger than the frequency obtained from the best-fit model of the pre-gap
TOAs at the same epoch (see Table~{\ref{table2}}). This, in principle, could
indicate that a glitch occured at some time during the gap. At MJD~51250,
midway between the pre-gap and the post-gap ephemerides, the fractional
change in frequency due to the possible glitch is
$\Delta$$\nu$/$\nu$~=~(7.11$\pm$0.15)$\times$10$^{-7}$. However, by using
six frequency derivatives, we found a possible ephemeris that fits the
entire Cycle~1 to~10 range (MJDs~50170 to~53787, see
Table~{\ref{table2}}). The RMS phase residual for that ephemeris is 0.019
(see Fig.~{\ref{figure4}}). Note that when finding an ephemeris that
spans several years, it is not uncommon to require a large number of frequency
derivatives in order to reduce the RMS phase residuals to a number on the
order of 5\%. This is typical especially in young pulsars and is generally
attributed to timing noise (e.g. \citealt{1509maggie}).
%% ======
%% |    |
%% |    |  FIGURE REZNOFIT.
%% |    |
%% ======
%% ======
%% |    |
%% |    |  FIGURE REZALL.
%% |    |
%% ======
%% ==
\par
%% ==
%% PARAGRAPH GLITCH IWASAWA.
The possibility of a glitch in 4U~0142+61 during our gap was in fact
examined by \citet{0142asca}. The authors showed that the frequency
obtained from a 1998 August {\em{ASCA}} observation of 4U~0142+61
(MJD~51046.699; 154 days after the start of our gap), differs from
the frequency predicted at the epoch of the observation by the ephemerides
previously reported for 4U~0142+61 in \citet{0142previous}.
\citeauthor{0142asca}, reported a frequency $f$~=~0.1150972(6)~Hz at
MJD~51046.69875 for the {\em{ASCA}} observation. Our overall ephemeris (see
Table~{\ref{table2}}) predicts $f$~=~0.115098404(3)~Hz at the same epoch.
Their measurement is within 2$\sigma$ of our prediction, indicating a
$\sim$5\% possibility that the two values are the same. Therefore, their
measured $f$ can be explained by a gradual change of the spin-down rate
without invoking a glitch.
%% ==
\par
%% ==
%% PARAGRAPH GLITCH INVESTIGATION.
However, the existence of our overall ephemeris cannot rule out the
possibility of the glitch: in some rotational glitches, the frequency
evolution, given some relaxation time after the glitch epoch, returns to
what it was prior to the glitch (see for example the glitch reported in
\citealt{1708glitch2}). If a glitch of this kind had happened inside the
two-year gap, and if the length of the gap was much greater than the
relaxation time, the only long-term effect of the glitch that could still be
observable with a timing analysis would be a random phase jump in the
post-relaxation TOAs relative to the pre-glitch TOAs. To investigate this
possibility, we added an arbitrary but constant time jump to all the
post-gap TOAs.  We were still able to find a new ephemeris that connected
the TOAs through the two-year gap. This indicates that our overall ephemeris
is not unique. Hence, we cannot rule out the possibility of a random phase
jump between Cycles~3 and~5, and therefore, a glitch cannot be ruled out.
%% ==
\par
%% ==
%% PARAGRAPH CAVEAT.
It is important to note that our method for obtaining TOAs,
(cross-correlating the folded profiles of given observations with a high
signal-to-noise template obtained from all the observations combined),
assumes a constant pulse profile. In the next Section, we show that the
pulse profile is actually varying during our monitoring program. However,
we performed simulations which showed that these changes do not result in
timing offsets significantly larger than the reported TOA uncertainties.
Hence the profile variations do not affect the above analysis.
%% --------------------------------------------------------
\subsection{Pulse Profile Changes}
\label{sec:profile}
%% --------------------------------------------------------
\subsubsection{Qualitative Observations}
\label{sec:qual}
%% --------------------------------------------------------
%% PARAGRAPH WITHOUT FSELECT.
%%%%%% {NOTE: Bla\verb|_|bla. check\_check. \verb|ftool_name|}.
We performed a pulse profile analysis using FTOOLS version
5.3.1\footnote{http://heasarc.gsfc.nasa.gov/ftools}. We used the following
steps: for each observation, we ran the FTOOL \verb|make_se| to combine the
GoodXenon files.  We then used the FTOOL \verb|fasebin| to make a
phase-resolved spectrum of the entire observation with 64 phase bins across
the profile. When we ran \verb|fasebin|, we selected layer~1 of the
detector, disregarded the propane photons, and included the photons from
PCUS~1,~2,~3, and~4. We disregarded photons from PCU~0 because of the loss
of its propane layer in 2000 \citep{propane} and because it gave different
results from the other PCUs. \verb|fasebin| also took care of barycentering
the data. For each observation, we then used \verb|seextrct| to make a
phase-averaged spectrum for the same set of detector layers and PCUs. The
phase-averaged spectrum was then used by the perl script \verb|pcarsp| to
make a response matrix.
%% ==
\par
%% ==
%% PARAGRAPH XSPEC STEP.
We loaded the phase-resolved spectra and the response matrices into the
X-ray Spectral Fitting Package (XSPEC\footnote{http://xspec.gsfc.nasa.gov
Version: 11.3.1}) and selected photons belonging to three energy bands:
2$-$10, 2$-$4, and~6$-$8~keV. Using XSPEC, we extracted an ASCII count rate
pulse profile for each of the energy bands. The profiles included
XSPEC-obtained 1$\sigma$ error bars on each of the phase bins in the
profiles. To obtain a pulse profile in units of count rate per PCU, we
divided the overall profile by a PCU coverage factor that took into account
the amount of time each PCU was on.  
%% ==
\par
%% ==
%% PARAGRAPH PROFILE TREATMENT.
We then aligned the 64-bin profiles with a high signal-to-noise template
using a similar cross-correlation procedure to the one described in the
timing analysis.  Then, for each {\emph{RXTE}} Cycle, we summed the aligned
profiles, extracted the DC component from the summed profile, and scaled the
resulting profile so that the value of the highest bin is unity and the
lowest point is zero. 
%% ==
\par
\begin{figure}
\includegraphics[scale=.72]{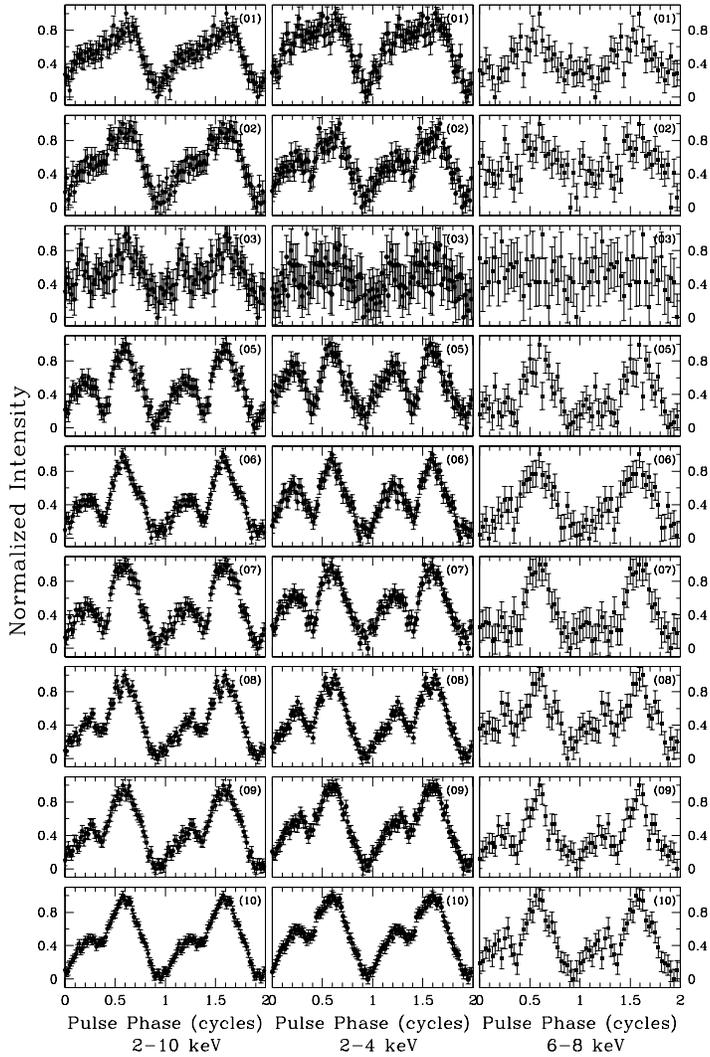}
\caption { 
Pulse profiles in all {\em{RXTE}} Cycles in three different PCA energy bands
(2$-$10~keV with 64 bins across the profile, 2$-$4~keV with 64 bins across
the profile, and 6$-$8~keV with 32 bins across the profile for clarity).
Note that no observations were made in Cycle~4. Two full periods are shown.
The normalization is such that the values of the lowest and highest bins
in each profile are 0 and 1, respectively. The Cycle number is shown in the
top right corner of each pulse profile plot. 
\label{figure5} 
}
\end{figure}

%% ==
%% PARAGRAPH PRESENTING PROFILES.
The average profiles in all three bands are presented in Figure~\ref{figure5}
for comparison. In a given band, the different profile qualities are due to
different net exposure times.
%% ==
\par
%% ==
%% PARAGRAPH QUALITATIVE SPECTRUM STATEMENT.
It is important to note that the two narrow energy bands that we are using
contain photons belonging to different spectral components: from the
spectrum of 4U~0142+61 (see, for example, \citealt{0142spectrum}), under the
assumption that the spectrum is well described by a blackbody plus power-law
tail, we know that the higher energy band (6$-$8~keV) contains negligible
blackbody emission, while the lower energy band (2$-$4~keV) contains
comparable amounts of blackbody and power-law emission. 
%% ==
\par
%% ==
%% COMMENT 1: DIP EVOLUTION, SECOND PEAK SOFTER.
Qualitatively, the evolution of the pulse profiles in the first two bands of
Figure~\ref{figure5} is clear to the eye. In Cycles~1 and~2, the smaller
peak, obvious in later Cycles, is not very well defined. After the two-year
gap, in Cycle~5, the `dip' between the peaks is much more pronounced. The
emission in the dip starts to most noticeably rise in Cycles subsequent to
Cycle~7. In the 6$-$8~keV band, the smaller peak appears to have lower
amplitude in the normalized profiles than in 2$-$4~keV, indicating that it
has a softer spectrum relative to the larger peak.
%% ==
\par
%% ==
%% COMMENT 2: RATIO OF HEIGHTS.
Another qualitative observation is that the ratio of the heights of the two
peaks in the 2$-$10~keV and 2$-$4~keV bands appears to be closest to unity
in the first Cycle after the gap. The ratio starts to decrease in the
Cycles subsequent to Cycle~5. Note that from this Figure alone we can
compare the sizes of the two peaks, but we cannot track the evolution of the
heights of each peak separately. In order to do that, we need to scale the
pulse profile of each Cycle by the average pulsed flux. This analysis is
presented in Section~\ref{sec:combo}.
%% ======
%% |    |
%% |    |  FIGURE PROFILES.
%% |    |
%% ======
%% --------------------------------------------------------
\subsubsection{Fourier Analysis}
\label{sec:fourier}
\begin{figure}
\includegraphics[scale=.82]{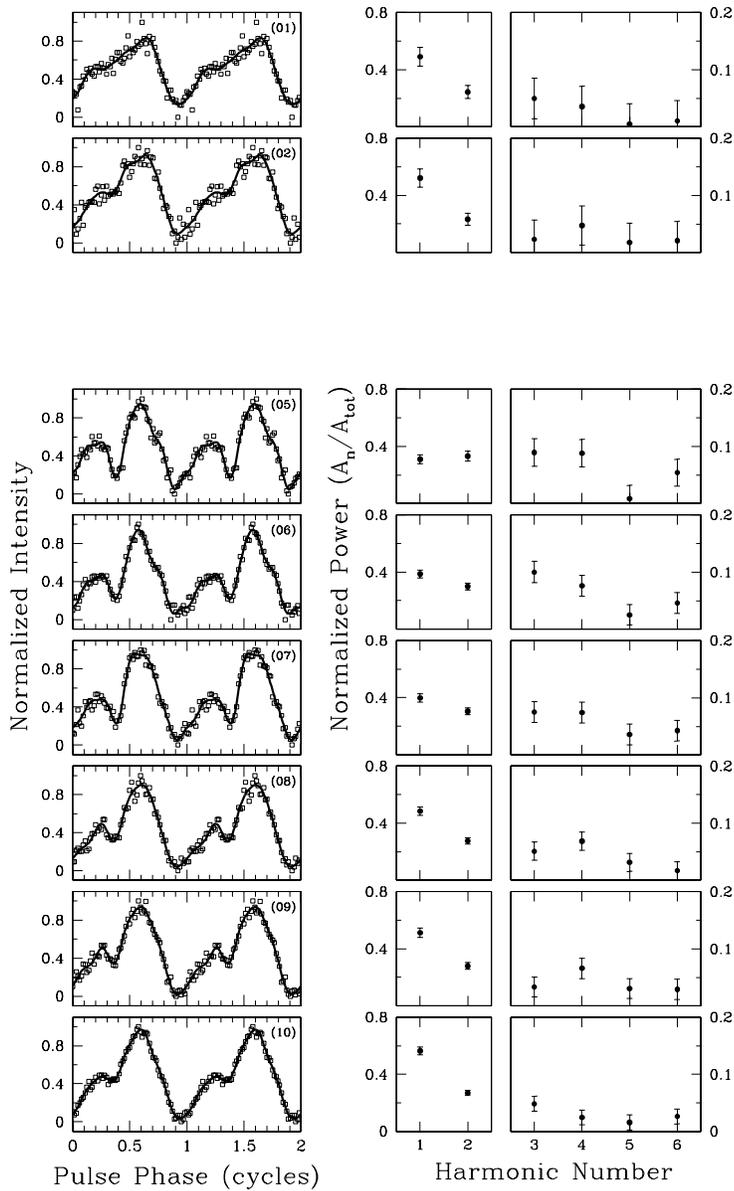}
\caption 
{ 
Fourier analysis of the pulse profiles in the 2$-$10~keV~energy 
band. Left: Pulse profile curves made of the first six calculated
Fourier components in each {\em{RXTE}} Cycle, superposed on the measured 
pulse profile points for that Cycle. Right: 
Harmonic content of the average pulse profiles for each Cycle.
Note that Cycle~3 was not included because of the poor signal-to-noise ratio.
\label{figurefou1} 
}
\end{figure}

\begin{figure}
\includegraphics[scale=.82]{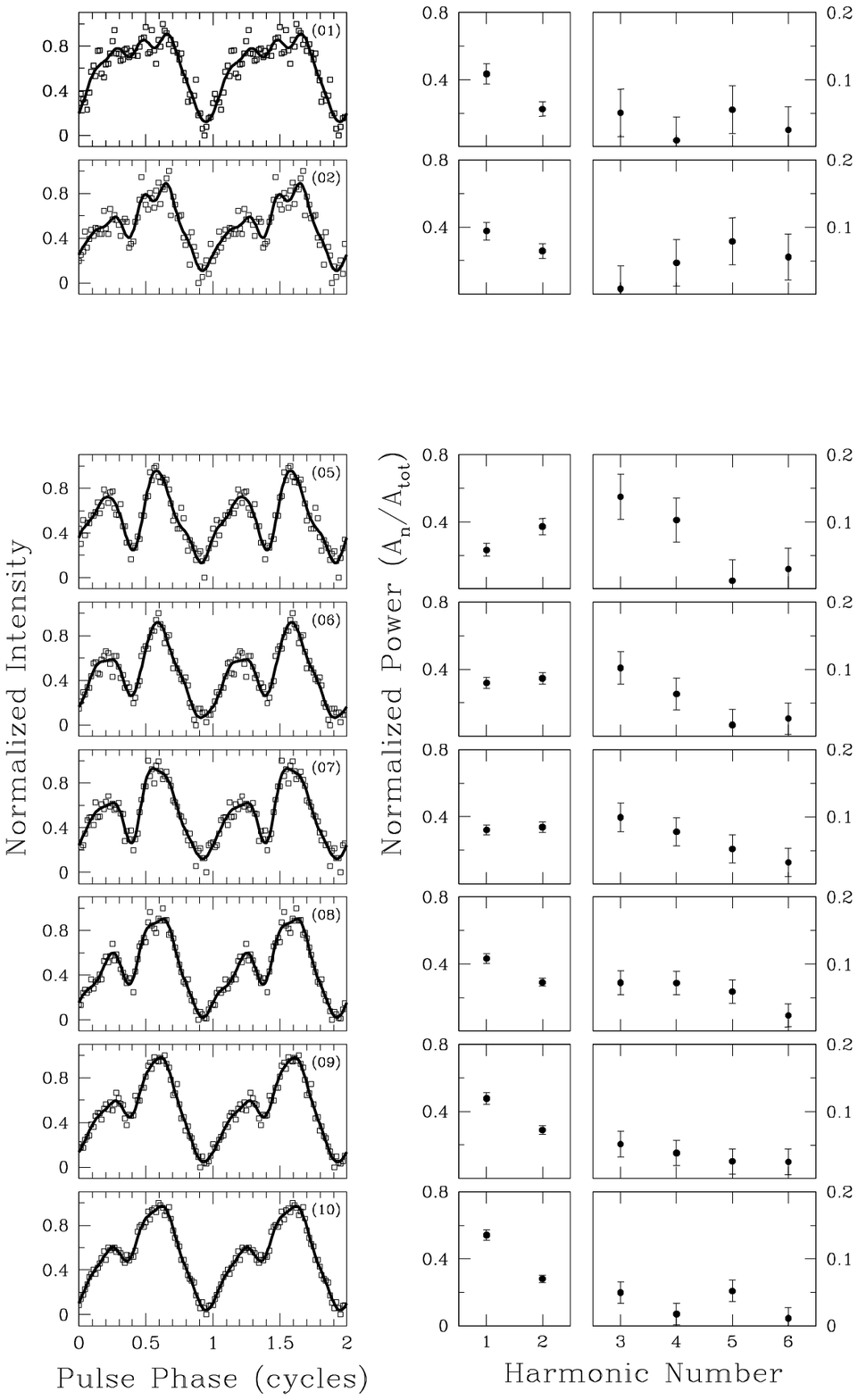}
\caption
{
Same as Figure~\ref{figurefou1} but for 2$-$4~keV.
\label{figurefou2}
}
\end{figure}

\begin{figure}
\includegraphics[scale=.82]{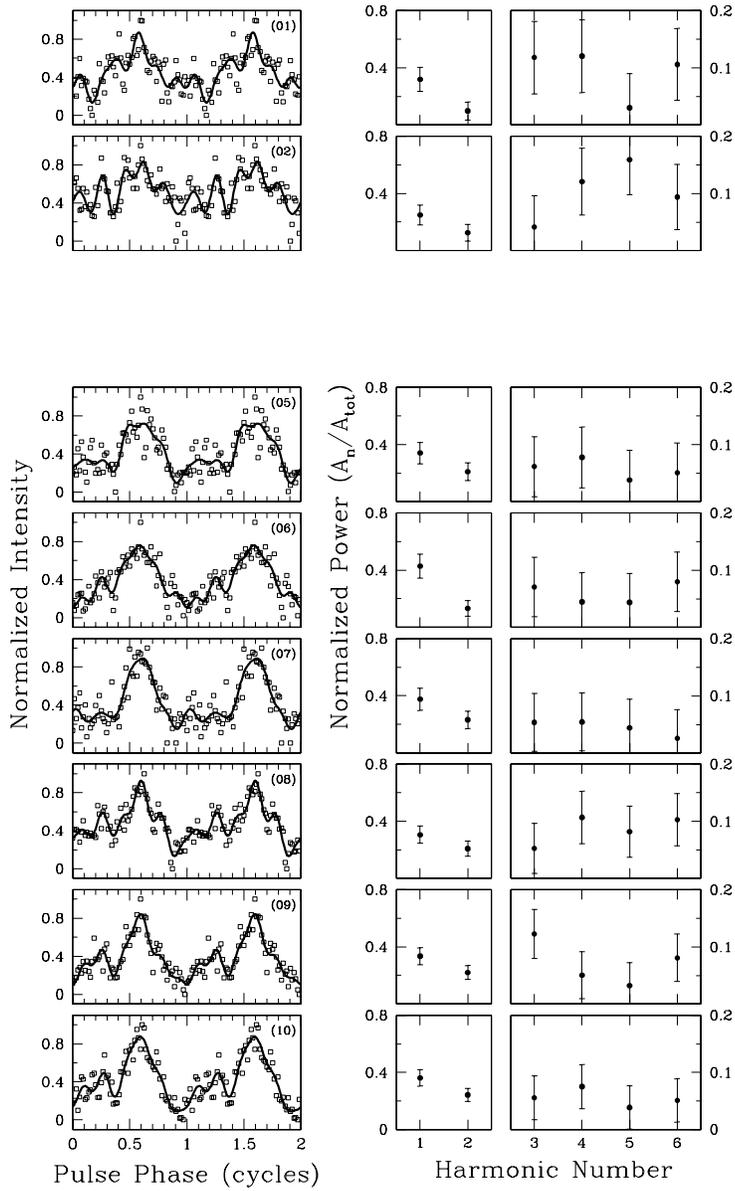}
\caption
{
Same as Figure~\ref{figurefou1} but for 6$-$8~keV.
%%%%%% unmentioned back to 64 bins.
\label{figurefou3}
}
\end{figure}

\begin{figure}
\includegraphics[scale=.8]{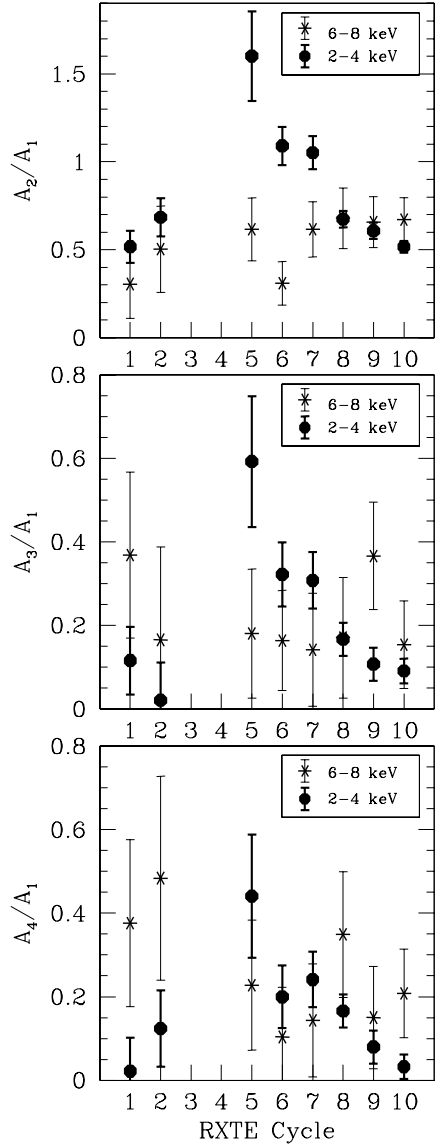}
\caption
{
Ratios of the Fourier amplitudes of the pulse profiles in two energy
bands. Top: ratio of the Fourier amplitude of the second harmonic to that of
the fundamental. Middle: ratio of the Fourier amplitude of the third
harmonic to
that of the fundamental. Bottom: ratio of the Fourier amplitude of the fourth
harmonic to that of the fundamental.
\label{figureratios}
}
\end{figure}

%% --------------------------------------------------------
%% PARAGRAPH INTRODUCING THE FIGURES.
To quantify the changes in the pulse profile, we computed the first six
Fourier amplitudes of the average profiles of each Cycle in each energy
band. Harmonic numbers larger than 6 were always consistent with zero. The
results are shown in Figures~\ref{figurefou1}, \ref{figurefou2}, and
\ref{figurefou3}. In each of the three Figures, the plots on the right show
the power in each harmonic divided by the total power in all harmonics (not
including the DC term). In the plots on the left, the observed pulse
profiles are in the background (hollow squares without error bars). The
superimposed smooth curve in the foreground is made from the first six
calculated Fourier components.
%%%%%% {without error bars}.
%%%%%% {NOTE THAT THE VALUES OF THE FOURIER COMPONENTS REMAINED THE SAME 
%%%%%% (WITIN ERRORS) WHEN THE PRE-GAP EPHEMERIS WAS USED IN CYCLES~1 AND~2 
%%%%%% AND THE SAME ANALYSIS WAS REPEATED, FIND A WAY TO SAY THAT HERE}.
The ratios of the first three Fourier harmonics relative to the fundamental
are presented in Figure~\ref{figureratios}. Note that the ratios of the
Fourier harmonics are not presented for Cycle~3 in Figures~\ref{figurefou1},
\ref{figurefou2}, and \ref{figurefou3} due to the very low signal-to-noise
ratio.
%% ======
%% |    |
%% |    |  FIGURE FOURIER WIDE.
%% |    |
%% ======
%%  ==
\par
%% ==
%% PARAGRAPH COMMENTS ON THE WIDE BAND FOURIER.
Significant variations were seen in the pulse profile of 4U~0142+61 in the
2$-$10~keV band and in 2$-$4~keV; on the left side of both
Figures~\ref{figurefou1} and~\ref{figurefou2}, the most striking variable
feature is the difference in the relative heights between the top of either
peak and the bottom of the dip. On the right side, considering the first two
Fourier components, the ratio of the second to first amplitudes
(A$_2$/A$_1$) is significantly bigger than unity only in Cycle~5. It then
falls steadily until, in Cycle~10, it reaches the same ratio as in Cycle~1.
The evolution of the A$_2$/A$_1$ ratio in 2$-$4~keV is shown in
Figure~\ref{figureratios}.  In the pre-gap Cycles, harmonics of order higher
than 2 are only marginally significant. In the post-gap Cycles, harmonics~3
and~4 are most significant in Cycles~5 and~6, coinciding with the
Cycles where the dip in the time-domain curve is sharpest. The
evolution of the A$_3$/A$_1$ and the A$_4$/A$_1$ ratios is also shown in
Figure~\ref{figureratios}. Note the obvious rise in the harmonic ratios just
post-gap in the soft-band, with apparent subsequent evolution to pre-gap
values.
%% ==
\par
%% ==
%% PARAGRAPH FOURIER 68.
In the 6$-$8~keV band (see Fig.~\ref{figurefou3}), statements about the
behavior of the Fourier components are harder to make because of the poor
signal-to-noise ratio; nevertheless some trends are clear.  Unlike in the lower
energy band, the A$_2$/A$_1$ ratio does not appear to systematically
increase or decrease.  Also unlike in the lower band, harmonics~3 and~4 do
not appear to vary systematically (see Fig.~\ref{figureratios}). Thus, in
the band where all the emission is from the power-law component of the
spectrum, the variations in the shape of the pulse profile, if any, are much
less significant than the variation in the 2$-$4~keV band which contains
photons belonging to both components of the spectrum. 
%% ======
%% |    |
%% |    |  FIGURE FOURIER24.
%% |    |
%% ======
%% ======
%% |    |
%% |    |  FIGURE FOURIER68.
%% |    |
%% ======
%% ======
%% |    |
%% |    |  FIGURE RATIOS.
%% |    |
%% ======
%% --------------------------------------------------------
\subsection{Pulsed Flux Time Series}
\label{sec:flux}
%% --------------------------------------------------------
%% PARAGRAPH TIME FORMULA.
%% PARAGRAPH METHOD 1.
To obtain a pulsed flux time series for 4U~0142+61, we did the following.
First, for each PCU in each observation, we used a procedure similar to that
described in Section~\ref{sec:qual} to make a phase resolved spectrum (with
16 phase bins across the profile) and a response matrix. We then used the
FTOOL \verb|fmodtab| to correct the exposure value in the phase resolved
spectrum of each PCU in order to take into account the amount of time that
each PCU was on. Then, for each observation, we added the spectra obtained
from PCUs~1 to~4 using the FTOOL \verb|fbadd| and added the responses using
the FTOOL \verb|addrmf|. We used \verb|fbadd| and \verb|addrmf| again to add
the spectra and responses of all observations in a given {\em{RXTE}} Cycle.
For each {\em{RXTE}} Cycle, we loaded the phase resolved spectra into XSPEC,
and selected photons in the 2$-$10~keV range. Using XSPEC, we extracted an
ASCII count rate pulse profile for each {\em{RXTE}} Cycle. The profiles
included XSPEC-determined 1$\sigma$ error bars on each of the phase bins. We
then smoothed each of the profiles by eliminating the Fourier components
corresponding to harmonic numbers larger than five. The pulsed flux for each
of the smoothed profiles was calculated using the following discrete area
formula:
%%%%
\begin{equation}
F = {\sum_{i=1}^{N}} ({p_i}-{p_{min}}) .
\end{equation}
%%%%
where $i$ refers to the phase bin, $N$=16 is the total number of phase bins,
$p_i$ is the count rate in the $i^{\textrm{th}}$ phase bin of the smoothed
pulse profile, and $p_{min}$ is the value of the minimum of the continuous
smooth function that is made from the first five Fourier components of the
original profile.
%% ==
\par
\begin{figure}
\includegraphics[scale=.60]{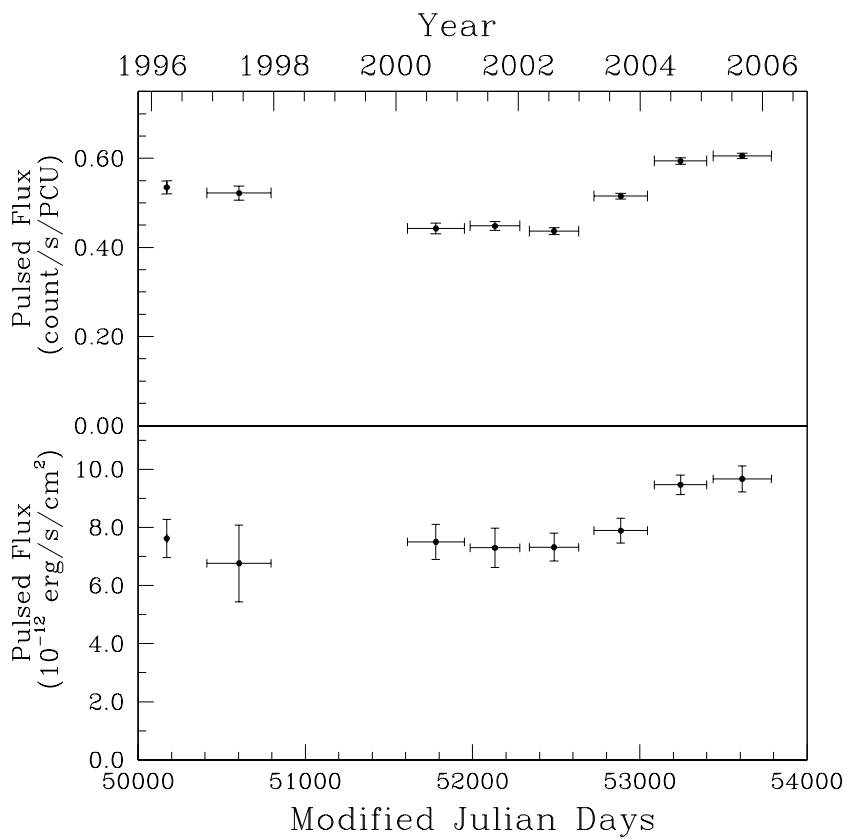}
\caption
{
%%%%%% =x=x=x=x=x=x
%%%%%% FIT PARAMETERS.
%%%%%% =x=x=x=x=x=x=
Top: Pulsed flux evolution of 4U~0142+61 in counts/s/PCU in the 2$-$10~keV
band. Each point corresponds to a full RXTE Cycle. Bottom: Pulsed flux
evolution in erg/s/cm$^2$ in the 2$-$10~keV band. See text for details.
\label{figure7} }
\end{figure}

%% ==
%% PARAGRAPH RESULTS.
The resulting pulsed flux history in counts/s/PCU is shown in the top panel
of Figure~\ref{figure7}. Each point represents one {\em{RXTE}} Cycle. The
pulsed flux has increased by 36$\pm$3\% between Cycles~7 and~9. A quick
rebinning of the observations shows that the increase period lasted
$\sim$~2.6~yr (between MJDs~52400~ and~53350). We verified that the same
trend is detected in PCUs~1$-$4 individually, and that there are no
comparable trends in the long-term light curves of the other 4~AXPs observed
as part of this monitoring program.
%% ==
\par
%% ==
%% PARAGRAPH MORE BANDS.
We repeated the above procedure of finding the flux for narrower energy
bands. There are hints that the long-term increase is present in the
2$-$4~keV band and not in 6$-$8~keV but our statistics do not let us confirm
this. If the pulsed flux increase is restricted to $<$~6~keV, this could
indicate that the spectrum of the pulsed emission is getting softer.
Motivated by this possibility, we performed a detailed spectral analysis of
four available archival {\em{XMM}} observations. We found that the spectrum
is indeed getting softer (Gonzalez~et~al. in preparation).
%% ==
\par
%% ==
%% PARAGRAPH NOTE ABOUT RMS.
Note that the method that we used to calculate the pulsed flux, which
consisted of calculating the pulsed area under the profile, is more
sensitive to noise than are measurements of the RMS pulsed flux like those
used in \cite{2259outburstw}. Therefore, to reduce the effects of
noise, it was necessary to combine the data from entire Cycles in order to
obtain each of the pulsed area points reported in the top panel of
Figure~\ref{figure7}, hence the large horizontal error bars. We
report measurements of the pulsed area instead of RMS
pulsed flux because, while it is true that measurements of the latter
are less sensitive to noise, changes in it can be
caused by changes in the real pulsed area and {\em{also}} by variations in
the pulse profile. As a
double check, we calculated the RMS pulsed flux for each observation, found
the average of the resulting fluxes in each {\em{RXTE}} Cycle, and multiplied
each average by a conversion factor dependent on the pulse shape, in order
to obtain a pulsed area. When we plotted these pulsed areas versus time, we
obtained a similar trend to that see in the top panel of
Figure~\ref{figure7}. For further discussion of the various methods used to
estimate the pulsed flux see Archibald et al. (2007, in preparation).
\par
%%
%% PARAGRAPH SPECTRAL FITTINGS.
In order to verify that the trend seen in the top panel of
Figure~\ref{figure7} is not an artifact of the response of the detector, and
in order to verify that the relative levels in the pre-gap and the post-gap
flux values are not skewed due to an evolution of the response of the
detector, we calculated the pulsed flux in erg/s/cm$^2$ using an additional
method that took the evolution of the response into account. For each
{\em{RXTE}} Cycle, we obtained one phase resolved spectrum (with 16 phase
bins across the profile) and one response matrix using the method described
above. This time we included PCU~0. Then, we defined the background phase
bin to be that where $p_{min}$ lies.  We then used the FTOOL \verb|cmppha|
to combine the spectra of all phase bins into a single phase-averaged
spectrum. Then, using XSPEC, we subtracted the spectrum of the background
bin from the combined spectrum of the remaining phase bins, scaling the
exposure appropriately. When subtracting the spectrum of the background bin
from the spectrum of the remaining bins, we assumed that the resulting
spectrum is, to a good approximation, that of the pulsed component of the
emission. For each {\em{RXTE}} Cycle, we fit the spectrum of the pulsed
emission with a model that consisted of a black body plus a power law. When
doing the fitting we froze the temperature of the black body to the value
$kT$~=~0.44~keV, the power-law photon index to $\gamma$~=~3.94 and the
column density of neutral hydrogen to
$N_h$~=~0.99$\times$10$^{22}$~cm$^{-2}$ . These parameters were obtained
from a linked spectral fit of the four archival XMM observations (see
Gonzalez et al. in preparation). We let the normalizations of the two
components of the spectrum vary freely. After the fit was done, we extracted
the pulsed flux numbers from the fitted spectrum of the pulsed emission of
each {\em{RXTE}} Cycle. We multiplied the 1$\sigma$ error bars returned by
XSPEC by the square root of the reduced chi squared of the spectral fit.
This multiplication is equivalent to assuming that the model being fit is
the right one to use and that any bad-fit results from an initial
underestimation of the error bars of the spectrum that is being fitted. The
resulting pulsed flux values in erg/s/cm$^2$ for each of the {\em{RXTE}}
Cycles are presented in the bottom panel of Figure~\ref{figure7}.  The trend
observed in the pulsed flux is similar to that in the top panel.  The
observed increase between Cycles~7 and~9 is 29$\pm$8\%, consistent with that
found in the first analysis. However, after having taken the response of the
detector into account, the pulsed flux in the pre-gap observations appears
to be consistent with the pulsed flux in the first post-gap Cycle.
%% ==
\par
%% ==
%% PARAGRAPH INCREASE ENERGETICS.
From the bottom panel of Figure~\ref{figure7}, the increase in the pulsed
flux is $\sim$~2.1$\times$$10^{-13}$~erg~s$^{-1}$~cm$^{-2}$ in the 2$-$10~keV band.
Assuming a distance of 2.5~kpc \citep*{0142ir}, the total luminosity
increase in the $\sim$~2.6-yr period during which the increase happened is
$\sim$~1.1$\times$$10^{33}$~erg~s$^{-1}$. This increase is of the same order
of magnitude as the average energy release rate in the first 1E~1048$-$5937
flare \citep{1048flares}.  It is also an order of magnitude smaller than the
average energy release rate in the first day following the outburst in
1E~2259+586 \citep{2259outburstw}.  The amount of energy released in the
same 2.6-year period due to the increase in the pulsed flux is
$\sim$~9$\times$$10^{40}$~erg in the 2$-$10~keV band. This is comparable to the
energy released in the second 1E~1048$-$5937 flare \citep{1048flares}, and
to the energy released in the in the year following the outburst in
1E~2259+586 \citep{2259outburstw}.
%% --------------------------------------------------------
\subsection{Combined Pulse Morphology and Pulsed Flux Analysis}
\label{sec:combo}
\begin{figure}
\includegraphics[scale=.72]{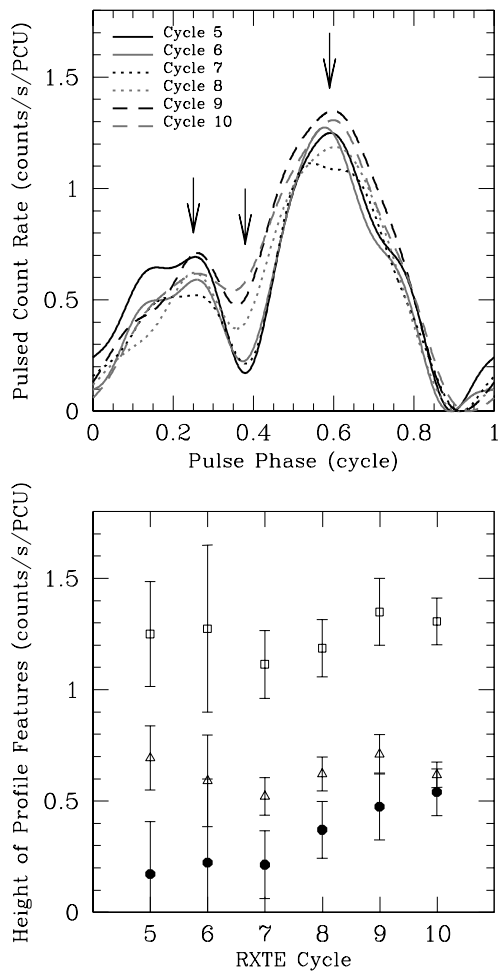}
\caption
{
Top: Superposed post-gap average pulse profiles in 2$-$10~keV
(with six Fourier components included),
scaled to give the appropriate average pulsed flux
for each {\em{RXTE}} Cycle. Bottom: 
The evolution of the heights of three different features
in the pulse profiles as a function of {\em{RXTE}} Cycle. 
The open triangles represent the maximum heights of the small peak
in each Cycle. The open squares represent the maximum heights of 
the big peak. The filled circles represent the 
heights of the dip. 
%%%%%% {CHECK HOW MUCH A CHANGE IN THE PULSED FLUX SLOPE WOULD
%%%%%% AFFECT THIS FIGURE}
%%%%%% {Uncorrected since the change from 19 to 24}.
\label{figurefeatures210}
}
\end{figure}

\begin{figure}
\includegraphics[scale=.72]{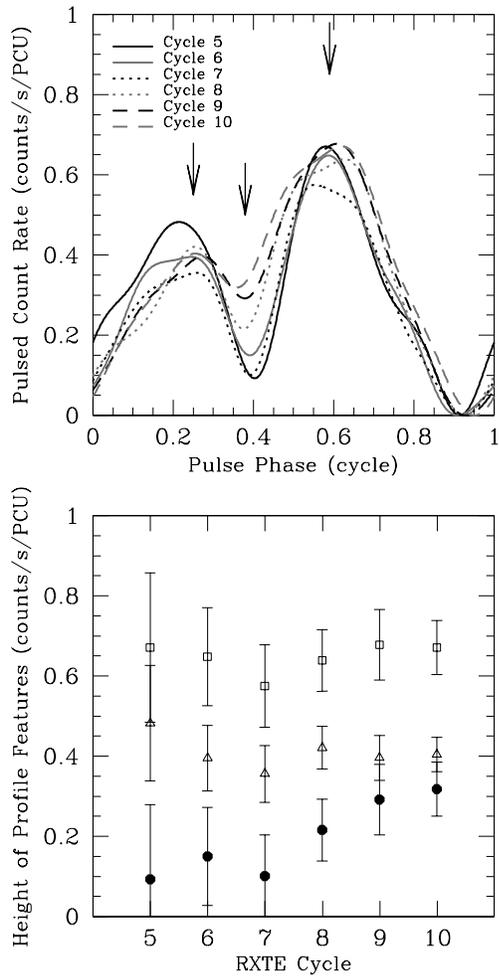}
\caption
{
See caption for Figure~\ref{figurefeatures210} but for 2$-$4~keV.
\label{figurefeatures24}
}
\end{figure}

%% --------------------------------------------------------
%% PARAGRAPH INTRODUCING FIGURES
In Section~\ref{sec:fourier}, we calculated the Fourier components of the
average pulse profiles. This gave us the relative amplitude of the pulse
profile harmonics in each {\em{RXTE}} Cycle. In Section~\ref{sec:flux}, we
calculated the pulsed flux for every observation. Here, we compute a
weighted average of the pulsed flux for each Cycle using the flux points
calculated in Section~\ref{sec:flux}. We then reconstruct the profiles for
each of the Cycles from the first six Fourier components (not including the
DC), scale them by the average RMS pulsed flux for that Cycle, and add the
necessary offset for the lowest point on each curve to be zero. This means
that the resulting scaled profiles return the correct pulsed flux.  The
advantage of this analysis is that we can now trace the evolution of each of
the peaks independently. The post-gap scaled profiles in 2$-$10~keV and in
2$-$4~keV are presented in the top panels of Figures~\ref{figurefeatures210}
and~\ref{figurefeatures24}, respectively. We did not include a similar
Figure for 6$-$8~keV because of the poor signal-to-noise ratio in that band.
The absolute heights of the peaks in the post-gap Cycles, as well as the
absolute height of the dip in between, are plotted in the bottom panels of
Figures~\ref{figurefeatures210} and~\ref{figurefeatures24}. The error bars
take into account both the errors on the Fourier components and the errors
on the pulsed flux.
%% ======
%% |    |
%% |    |  FIGURE  MIDCURVES 1.
%% |    |
%% ======
%% ======
%% |    |
%% |    |  FIGURE  MIDCURVES 2.
%% |    |
%% ======
%% ==
\par
%% ==
%% PARAGRAPH REMINDER.
In both Figures, there is a hint of increase in the height of the big peak 
between Cycles~7 and~9. The dip between the peaks appears to be getting
shallower more rapidly between Cycles~7 and~9.
%%%%%% In both Figures, the height of the small peak is slowly increasing. The
%%%%%% height of the big peak is increasing more rapidly than that of the small
%%%%%% peak. The dip between the peaks appears to be getting shallower rapidly also.  
The difference in the height of the dip over these 3~years is more
significant than the difference in the height of either peaks.  This
indicates that the biggest contribution to the change in the pulsed flux
comes from an increase in the emission in the dip, which, in principle,
could be caused by the widening of either peak around the dip.
%% --------------------------------------------------------
\section{Discussion}
\label{sec:discussion}
%% --------------------------------------------------------
\subsection{Possible Event in the Gap?}
\label{sec:eventgap}
%% --------------------------------------------------------
%% PARAGRAPH EVENT1.
Could a short-time scale energetic event (such as an outburst like that seen
in 2002 for 1E~2259+586) have occured sometime within the two-year gap and
triggered the pulsed flux and pulse profile changes that we are observing?
As discussed above, the possibility of a glitch during the gap was examined
by \citet{0142asca}. Here, an examination of our timing, flux, and pulse
profile analyses can provide further clues to help answer this question.
%% ==
\par
%% ==
%% PARAGRAPH EVENT 2.
From our timing analysis (Section~\ref{sec:timing}), there is some evidence
for a glitch having occured sometime during our gap. Hence, if an outburst
did occur, it might have been accompanied by a glitch, as was the case for
the 2002 outburst of 1E~2259+586 \citep{2259outburstk}.  If there was a
glitch in 4U~0142+61, the unrecovered fractional change in frequency would
have been (7.11$\pm$0.15)$\times$10$^{-7}$, a factor of~6 smaller than the
maximum fractional frequency change of (4.24$\pm$0.11)$\times$10$^{-6}$
observed in 1E~2259+586. The fact that the pulsed flux in the first post-gap
observations is consistent with that in the last pre-gap observations could
be consistent with an outburst in between if the initial flux increase
during an outburst had time to die down (see Fig.~\ref{figure7}). If we
assume that the return of the pulse profile to its pre-gap shape is a
recovery following an outburst, this would imply a much longer timescale for
the pulse profile relaxation phase than for the pulsed flux relaxation, the
opposite to what was seen following the 1E~2259+586 outburst
\citep{2259outburstw}.  
%%%%%%
Alternatively, the post-outburst pulse profile relaxation could have been
completed during the gap, and the slow return of the post-gap profile to its
pre-gap morphology could be attributed to a different phenomenon. This is
further discussed in Section~\ref{sec:phys}. In either case, if there was an
event during the gap, why the pulsed flux is presently rising is unclear. If
the event associated with the putative glitch released energy deep in the
neutron-star crust, then the increase could be due to its slow release (eg.
\citealt{thermal1}; \citealt{thermal2}). Given the size of the observed flux
increase and its timescale for release, the initial energy deposition would
have had to have been large, $\sim$10$^{45}$~erg \citep{thermal2}. This is
comparable to the observed energy release in giant SGR flares \citep{hbs}.
%% ==
\par
%% ==
%% PARAGRAPH ASCA.
If we assume that the pulse shape prior to the gap is the ``relaxed'' pulse
shape, the evolution of the harmonic ratios shown in
Figure~\ref{figureratios} supports the possibility of relaxation of the
profile following an event in the gap. To shed light on the events in the
gap, we can compare our {\it RXTE} profiles with those observed with
{\em{ASCA}} by \citet{0142asca}. In their Figure~4, pulse profiles in
0.5$-$10~keV for a) September 1994, b) August 1998, and c) combined July and
August 1999 profiles are presented. In the 1994 and 1998 observations, the
profile consisted of two peaks, with the trailing peak being the smallest,
with the dip between the peaks being higher than the lowest bin in the
profiles. The shape of the 1998 profile is in agreement with the {\em{RXTE}}
pre-gap average pulse profiles for Cycle~2 (see Fig.~\ref{figurefou1}). In
the 1999 {\em{ASCA}} profile, the amplitude of the trailing peak was
{\em{higher}} than that of the leading peak. In addition, the difference
between the height of the dip and the lowest point in the profile decreased.
Interestingly, the changes in the {\it ASCA} profiles appeared more
significant at the lower end of the energy band, as we observe in our
{\em{RXTE}} data.  In 2000, the first {\em{RXTE}} Cycle after the gap has a
profile in which the trailing peak is once again smaller than the leading
peak.
%%%%%% \footnote{Note that the fact that the two peaks have different spectra
%%%%%% provided an extra check that we did not accidentally swap the peaks during
%%%%%% the cross-correlation part of our analysis}.  
The dip between the peaks, however, is still more pronounced than in the
pre-gap observations.
%% ==
\par
%% ==
%% PARAGRAPH OVERALL.
Overall, the timing data and the pulse profile data are consistent with 
some sort of event, possibly a glitch with accompanying sudden pulse profile
change, having occured between 1998 August and 1999 July, possibly with
the latter's long-term relaxation still ongoing as of early 2006. However,
we suggest an alternate explanation for the latter point below.
%% --------------------------------------------------------
\subsection{Brief Review of the Magnetar Model}
\label{sec:theory}
%% --------------------------------------------------------
%% PARAGRAPH SCATTERING.
In the detailed magnetar model proposed by \citet{tlk02rim}, the crust of a
magnetar is deformed by internal magnetic stresses, thereby twisting the
footpoints of the external magnetic field, driving powerful currents in the
magnetosphere and twisting the magnetosphere relative to the standard
dipolar geometry.  These magnetospheric currents resonantly cyclotron
scatter seed surface thermal photons.  The seed contribution to the thermal
component of the spectrum is thought to arise from heat resulting from the
active decay of a high internal magnetic field \citep{td96rim,tlk02rim}. 
The magnetospheric scattering is responsible for the non-thermal component
of AXP spectra.  Additionally, the surface is back-heated by the currents,
resulting in additional thermal emission. Indeed, the persistent emission in
AXPs generally has a spectrum that is well described by a two-component
model, consisting of a blackbody plus a hard power-law tail, as expected in
this model.  
%% ==
\par
%% ==
%% PARAGRAPH STATEMENT, SOME CORRELATIONS ARE EXPECTED.
Changes in pulsed and/or total X-ray luminosity, spectral hardness, and
torque are predicted to have a common physical origin in the
\citet{tlk02rim} model and some correlations are expected. 
%%%%%% {NOTE1: THERE
%%%%%% REALLY ISN'T MUCH SAID ABOUT THE LUMINOSITY IN TLK02}. {NOTE2: IN THE PAPER
%%%%%% THAT SUMMARIZES TLK02 THERE IS A STATEMENT ABOUT THE PULSE FRACTIONS THAT
%%%%%% SEEMS TO COME OUT OF NOWHERE. WHERE IS THE PHYSICAL EXPLANATION FOR IT????}.
%%%%%% PAR. TWIST AND CURRENT DISTRIBUTION RELATED. LARGER TWISTS, HARDER SPECTRA.
Changes in twist angle of the magnetic field, cause, or may be caused by,
changes in the magnetospheric current distribution (due either to sudden
crustal deformation like in AXP~outbursts and SGR~giant flares or due to
slower crustal deformations as may be taking place in the AXP~flares;
\citealt{1048flares}). Larger twists generally correspond to harder
persistent X-ray spectra and higher magnitudes of the spin-down rates, as is
observed when comparing the harder SGR spectra to those of the softer AXPs
\citep{mw01rim}.  A similar trend might be expected for a single magnetar
exhibiting luminosity variations: a higher luminosity should correspond to a
larger twist, hence harder spectrum, as has been reported for
1RXS~J170849.0$-$400910 (\citealt{1708hardness}; \citealt{1708campana}).
A higher luminosity
should also in general correspond to a larger magnitude of the spin-down
rate. However, decoupling between the torque and the luminosity can
occur because the torque is most sensitive to the current flowing on a
relatively narrow bundle of field lines that are anchored close to the
magnetic pole.  For a single source, whether an X-ray luminosity change will
be accompanied by a torque change depends on where in relation to the
magnetic pole the source of the enhanced X-rays sits.
%%%%%% This is the explanation cited by~\citep{1048flares} because of the absence
%%%%%% of that correlation. But in hindsight, the overall correlation might be
%%%%%% there given that the pulsed flux was low when the spin-down was steady.
%% ==
\par
%% ==
%% PARAGRAPH PROFILE FEATURES.
The \citet{tlk02rim} model can also explain properties of the pulse profiles
of magnetars. According to the model, several effects can affect the pulse
shape, generate subpulses, and/or increase the energy dependence of the
pulse profile. 
%%%%%% These effects are: {\em{a)}} anisotropy in the angular
%%%%%% PARAGRAPH SIMPLIFICATION.
%%%%%% Note also that an increase in the twisting angle of the magnetic field
In addition, an increase in the twisting angle of the magnetic field
increases multiple scattering and increases the optical depth to resonant
scattering which can simplify the pulse shape. This is one of the two
proposed explanations for the sudden simplification of the pulse profile of
SGR~1900+14 after its dramatic giant-flare \citep{sgr1900changes}, the other
explanation being the sudden elimination of the nonaxisymmetric components
of the magnetospheric currents \citep{tlk02rim}. 
%% --------------------------------------------------------
\subsection{Possible Physical Interpretations for 4U~0142+61}
\label{sec:phys}
%% --------------------------------------------------------
%% PARAGRAPH PHYSSTART.
In this paper, we have shown that the pulsed flux of 4U~0142+61 has
increased on a time scale of a few years, and we have found simultaneous
slow pulse profile evolution in the 2$-$4~keV band, which may to be
recovering from some event that occured prior to 2000 but after 1997,
possibly in the interval between 1998 August and 1999 July. We have also
found evidence, as first suggested by \citet{0142asca}, that there was a
timing glitch in that same interval, although we cannot confirm its
existence.  Can the magnetar model explain these observations?
%% ==
\par
%% ==
%% PARAGRAPH EXPECT BOTH COMPONENTS TO VARRY.
The energy dependence of the pulse profile evolution is puzzling.  As
described above, in the twisted magnetosphere model of \citet{tlk02rim}, the
non-thermal emission in an AXP is the result of magnetospheric scattering of
surface thermal photons.  If the surface emission angular pattern were
changing, therefore, the non-thermal angular pattern should as well. That we
do not observe comparable pulse profile changes in the 6$-$8~keV band for
4U~0142+61 is thus puzzling.  One possibility is that the seed thermal
emission is not changing appreciably, but the scattering currents in the
outer regions of the magnetosphere, where the cyclotron energy is lower, are
changing, while the inner currents are not.  This could arise if there is
evolution in the field configuration closer to the magnetic poles, with
relatively little closer to the magnetic equator. Why variations in the field
configuration should be geographically localized, however, is unclear.
%% ==
\par
%% ==
%% PARAGRAPH THERMAL GLITCH.
The increase in the pulsed flux over a similar time scale as that of the
profile evolution is apparently accompanied by a softening of the spectrum
(Gonzalez~et~al. in preparation). As discussed above, the putative 1998/1999
glitch may have deposited a large amount of energy in the crust, with it
only starting to be radiated away in 2002. The energy released would have
had to have been large, $\sim$10$^{45}$~erg \citep{thermal2}. Such a thermal
energy release could be influencing the pulse profile as well, although some
change in profile would be expected in the hard band too, which is not
observed. Moreover, such an increase has not been observed following the
glitches in 1E~2259+586 or 1RXS 170849.0$-$400910, although this could be a
result of smaller total energy releases.
%% ==
\par
%% ==
%% PARAGRAPH MENTION TWO POSSIBILITIES.
Alternatively, the increase in pulsed flux seen in 4U~0142+61 between 2002
and 2005 could be explained by the twisted magnetosphere model. In this
framework, there are two possibilities: {\em{a)}} a slow {\em{in}}crease in
the twist of the magnetic field lines in the magnetosphere, or, {\em{b)}} a
slow {\em{de}}crease in the twist angle.
%% ==
\par
%% ==
%% PARAGRAPH FIRST POSSIBILITY.
In the first possibility, the observed increase in the pulsed flux could be
an extreme case of the 1E~1048$-$5937 flares, i.e. a slow twisting of the
magnetic field lines in the suggested interpretation of \citet{1048flares}.
This explanation is only valid if the total flux, which remains to be
determined with an imaging telescope, is increasing as well. How this would
be related to the putative 1998/1999 event is unclear; any flux enhancement
that occured then would have had to have largely decayed away by 2000. In
any case, for an increase in the twist angle, one expects a spectral
hardening \citep{tlk02rim}. We do not observe this. Also for an increase in
the twist angle, at least naively, the twisted magnetosphere model predicts
an increase in torque as the flux rises \citep{tlk02rim}. From
Table~\ref{table2}, the post-gap ephemeris $\ddot{\nu}$ is positive, meaning
$\dot{\nu}$ is increasing, i.e. the magnitude of the pulsar's spin down rate
is decreasing, the opposite of what is expected for a flux increase, unless
the magnetospheric currents causing the torque are flowing only in the small
polar cap region.  Finally, if the slow increase in pulsed flux were caused
by a slow magnetospheric twisting, the decrease in the size of the Fourier
components of order higher than unity in the low energy band could be
interpreted as a simplification of the pulse profile due to an increase in
the scattering in the magnetosphere.  However, the most extreme case of a
pulse profile simplification, which was reported in SGR~1900+14, happened
equally in all bands following a dramatic increase in the scattering after a
giant flare (\citealt{sgr1900changes}, \citealt{tlk02rim}). The phenomenon
that we have observed, by contrast, is restricted to the softer energies. 
%% ==
\par
%% ==
%% PARAGRAPH BUILDUP.
In the previous paragraph, we mentioned that the increase in the pulsed flux
and the gradual changes in the pulse profile may indicate stress build-up
caused by an increase in the twist angle in the magnetosphere. The pulse
profiles of Cycles~2 and~10 are very similar, indicating that the pulse
profile of Cycle~2 may also be showing signs of the same kind of stress
build-up. Under these assumptions, if an event occured in the gap following
Cycle~2, it is reasonable to expect a similar event to follow after
Cycle~10. Indeed, in 2006 April, less than 2~months after the end of
Cycle~10, the pulsar appears to have entered an extended active phase: a
single burst accompanied by a pulse profile change was detected from the
pulsar on April~6~\citep*{atel1}. A series of four bursts was later detected
on June~25~\citep{atel2} and a larger burst was detected on 2007
February~07~\citep{0142lastburst}. A detailed paper on these events is
currently in preparation. If they are indeed due to a stress release
following several years of slow magnetospheric twist, then the causes of the
softening of the spectrum and of the decrease in the magnitude of the
pulsar's spin-down are unclear.
%% ==
\par
%% ==
%% PARAGRAPH UNTWIST
The second possibility in the framework of the twisted magnetosphere model,
is that the observed increase in the pulsed flux is accompanying a slow
{\em{de}}crease in the twist angle of the magnetic field lines in the
magnetosphere. This explanation is only valid if the total flux is
decreasing. If the total flux were falling with the pulsed fraction rising,
the observed spectral softening could be consistent with naive
{\em{un}}twisting expectations as would the decreasing spin-down rate.
Indeed an anti-correlation between total flux and pulsed fraction was
observed by \citet{tiengo2005} for 1E~1048.1$-$5937, supporting this
possibility. However, the problem of why the pulse profile is evolving at low,
but not high energies, remains.
%%%%%% ==
%%%%%% \par
%%%%%% ==
%%%%%% PARAGRAPH SUMMARIZE PHYS.
%% --------------------------------------------------------
\subsection{Other Wavelengths}
\label{sec:other}
%% --------------------------------------------------------
%% PARAGRAPH MULTI-WAVELENGTH.
4U 0142+61 is truly a multi-wavelength AXP. It is known to pulsate in the
optical band \citep{0142opti1,0142opti2} and it has been detected in the
near-IR \citep{0142ir}, in the mid-IR using {\em{SPITZER}} \citep*{disk},
and in hard X-rays \citep{integralkuiper, integralhartog1}.
%% ==
\par
%% ==
%% PARAGRAPH ORIGIN.
The origin of the emission at these other wavelengths remains unclear, although
some models have been proposed. In hard X-rays, \citet{beloborodov} argue
that the emission is either due to bremsstrahlung photons emitted by a thin
surface layer, or is due to synchrotron emission originating from the region
in the magnetosphere where the electron cyclotron energy is in the keV
range. The near-IR and optical emission is thought to be magnetospheric
%%%%%% (\citealp*{irtheory1}; \citealt{irtheory2}), 
\citep*{irtheory1}, while the mid-IR emission is suggested to be due to a
passive fall-back disk \citep{disk}.
%% ==
\par
%% ==
%% PARAGRAPH CORRELATIONS.
Looking for correlations between the X-rays and the emission in other
wavelengths may serve as tests of emission models, as correlations between
X-ray and near-IR fluxes have sometimes been observed. For example, a
correlation in the decays of the X-ray and near-IR fluxes in 1E~2259+586 was
observed following the 2002 outburst \citep{2259ir}; a similar
correlation was reported for XTE~J1810$-$197 \citep{1810ir}. However, in
other instances, the two fluxes were not correlated, as for 1E~1048$-$5937
(\citealt{1048flares}, \citealt{1048unpulsedflares}, and
\citealp*{1048ir3}). In 4U~0142+61, reported variations in the IR are not
seen contemporaneously in our X-ray data, and are on time scales much
shorter than that of the X-ray variation reported here \citep{0142ir,
martinnh}.
%% ==
\par
%% ==
%% PARAGRAPH DISK 1.
\citet{disk} observed mid-IR emission from 4U~0142+61 which they argue is
associated with a passive fall-back disk irradiated by the central X-ray
pulsar. If this is the case, then if 4U~0142+61's X-ray flux is increasing,
one expects a corresponding increase in the disk emission.  It is thus
important to establish the behavior of the total flux, in addition to that
of the pulsed flux reported on here.
%% == 
\par
%% ==
%% PARAGRAPH DISK 2.
The possible presence of a disk suggests that if a sudden, impulsive
outburst occured in the gap, the energy released must have been
significantly smaller than the disk binding energy, after accounting
for the disk thickness. 
%%%%%% =X=X=X=X=X=X=X=X=
%%%%%% CALCULATION
%%%%%% inner radius = X solar radius
%%%%%% outer radius = X solar radii
%%%%%% mass = about X earth mass
%%%%%%
%%%%%% giant flare SGR 1900+14 energy release (1998, August)
%%%%%% *initial pulse 6.8E43
%%%%%% *tail          5.2E43
%%%%%% *total bigger than 1.2E44 erg
%%%%%% giant flare SGR 0526-66 (1979)
%%%%%% *initial pulse 1.6E44
%%%%%% *tail          3.6E44
%%%%%% *total         5.2E44
%%%%%% ref: {flareenergy}
%%%%%%
%%%%%% KEEPING TRACK OF OTHER AXP ENERGY NUMBERS:
%%%%%% 1048 numbers: 2.7E40 erg (flare1), 2.8E41 erg (flare2).
%%%%%% 2259 numbers: 2.7E39 erg (day1), 2.1E41 erg (year 1 - day 1).
%%%%%% SGR~1900+14 or the SGR~0526-66 gflares (lost the info: around E40).
%%%%%% =X=X=X=X=X=X=X=X=
For a central pulsar mass of $M_{psr}\sim$~1.4$\Msun$, a uniform disk of
mass $M\sim$~3$\Mearth$, and inner and outer radii $R_1\sim$~3$\Rsun$ and
$R_2\sim$~10$\Rsun$ \citep{disk} the disk binding energy is 
%%%% $G \frac{M_{psr} M}{R_1+R_2}$~$+$~$G M^2 \frac
%%%% { 2( {R_2}^3 +2 {R1}^3 -3 {R_1}^2 {R_2}) }{ 3( {R_2}^2 -{R_1}^2 )^2 }$
$\sim$4$\times$10$^{42}$~erg. Since the X-ray luminosity of the source
($\sim$10$^{35}$~erg/s) integrated over a period $>5000$~yr is much larger
than the binding energy of the observed disk, one must assume that when the
source is not undergoing an outburst, the disk is in an equilibrium state
where the rate of energy absorption is balanced by the rate of disk
emission. Assuming this equilibrium cannot hold on the timescale of a sudden
outburst, then IR observations of the disk provide an upper limit of
(4$\times$10$^{42}$/$f$)~erg on the energy released in a possible outburst
in the gap, where $f$ is the fraction of the solid angle occupied by the
thickness of the disk. For $f$~=~0.01, this upper limit is three orders of
magnitude larger than the total energy released during the flares of
1E~1048$-$5937 \citep{1048flares}. It is also five orders of magnitude
larger than the energy released during the first day of the 1E~2259+586
outburst \citep{2259outburstw}.  Thus, an event in the gap of either the
magnitude of the flares or that of the outburst occuring could have affected
the disk but seems unlikely to have disrupted it. For $f$~=~0.01, the upper
limit is also of the same magnitude as the energy released in either of the
SGR~1900+14 or the SGR~0526$-$66 giant flares \citep{flareenergy}. This
suggests that there have been no events of the magnitude of the giant SGR
flares since the putative disk's formation. Given that we have witnessed 3
giant SGR flares each from a different source since 1979, and none from
AXPs, this suggests that AXPs do not exhibit giant flares. This would
suggest an interesting distinction between SGRs and AXPs: if the two are
evolutionary linked, the AXP phase must come first, unless a debris disk can
reform following a giant flare.
%% == 
\par
%% ==
%% PARAGRAPH ACTIVE DISK.
The mid-IR emission has also been interpreted as an active fall-back disk in
which the pulsar is accreting in a propeller mode. In this case, the dipole
field strength of the pulsar is typical of conventional radio pulsars, ie.
$\sim~10^{12}$~G \citep{moredisk}, although the surface field strength is in
the magnetar range, owing to higher order multipoles.
%%%% In this case, the dipole field strengths of the pulsars
%%%% are {\it not} in the magnetar range, though the surface 
%%%% field strengths are, owing to magnetar-strength higher order multipoles.
%%%%
%%%% The mid-IR emission has also been interpreted as an active fall-back disk in
%%%% which the pulsar is accreting in a propeller mode, and has a surface dipole
%%%% magnetic field strength typical of conventional radio pulsars, ie. $\sim
%%%% 10^{12}$~G \citep{moredisk}. 
In this model, the X-rays arise from propeller accretion. Without detailed
models of the spectra and of the pulse shapes expected in this model, we
cannot interpret our observations in this framework.  Nevertheless, one
might expect a torque change with luminosity in this model.  For 4U~0142+61,
as noted in Section~\ref{sec:phys}, the magnitude of the torque is decreasing while
the pulsed flux is increasing. If propeller accretion is occuring, then the
total flux should be decreasing; this can be checked (Gonzalez et al., in
preparation). We note that evidence against such a torque/luminosity
correlation has been presented by \citep{1048flares} for a different AXP.
%%% Also, if the 2006 and 2007 bursting activity is related to the pulsed flux
%%% increase, this would be further evidence against an accretion disk model
%%% because it postulates that the bursting is from an independent mechanism
%%% (namely effects of magnetar strength higher order multipoles).
%% --------------------------------------------------------
\section{Summary}
\label{sec:sum} 
%% --------------------------------------------------------
%% PARAGRAPH SUM.
Our continuing {\em{RXTE}} monitoring program has revealed a possibly new
AXP variability phenomenon: 4U~0142+61 exhibited a slow but steady
increase in its pulsed flux between 2002 May and 2004 December, such that it
has risen 36$\pm$3\% over 2.6 years in the 2$-$10~keV band. This is
accompanied by a softening of the spectrum (Gonzalez~et~al. in preparation).
Quasi-simultaneously, the pulse profile, which
comprises two peaks having different spectra, has been evolving since 2000.
In particular, the dip of emission between the two peaks has been rising
since 2002, as if it is returning to its pre-2000 morphology in which there
was no clear distinction between the peaks. The profile evolution translates
to a reduction of the power in the Fourier harmonics of order higher than
one since 2000. This is in contrast with the pulsed flux which seems to be
moving {\em{away}} from the pre-2000 value. The evolution in the pulse
profile is seen in the 2$-$4~keV band but not in the 6$-$8~keV band,
presenting an interesting puzzle to the twisted magnetosphere model for
magnetars.  Intriguingly, \citet{0142asca} have suggested the pulsar
suffered a glitch just before 2000 on the basis of a single discrepant
{\em{ASCA}} period measurement. Our phase-coherent timing using {\em{RXTE}}
demonstrates that a glitch is plausible but not necessary to explain the
data, but our pulse profile evolution analysis provides new evidence for
such an event having occured.
%% ==
\par
%% ==
%% PARAGRAPH PHYS IN SUM.
Physical interpretations that described well other observed long-term
changes in AXP emission (such as outburst afterglow or flux flares caused by
an increased twist in the magnetosphere) do not explain all the phenomena
that we have observed.  Most of our observations could be explained by the
twisted magnetosphere model if the total flux of 4U~0142+61 is actually
decreasing. This would indicate a slow {\em{un}}twisting in the
magnetosphere.  Alternatively, if the total flux is increasing, a slow
{\em{increase}} in the twist angle in the magnetosphere can account for the
pulse profile simplification and for the post-Cycle~10 events, but the
changes in the spin-down rate and the softening of the spectrum would remain
unexplained.  Finally, the data could be explained by energy release
following an energy deposition, perhaps due to a glitch that occured in the
crust of the star sometime during the post Cycle~2 gap, although the energy
deposited would have had to have been large. No matter what, the absence of
profile evolution at high energies remains a puzzle.
%% --------------------------------------------------------
\acknowledgments
%% --------------------------------------------------------
We thank Andrew Cumming for helpful conversations. This work was supported
by the Natural Sciences and Engineering Research Council (NSERC) PGSD
scholarship to RD. FPG is supported by the NASA Postdoctoral Program
administered by Oak Ridge Associated Universities at NASA Goddard Space
Flight Center. Additional support was provided by NSERC Discovery Grant
Rgpin 228738-03, NSERC Steacie Supplement Smfsu 268264-03, FQRNT, CIAR, and
CFI.  VMK is a Lorne Trottier and Canada Research Chair.
%% --------------------------------------------------------
%% \bibliographystyle{apj}
%% \bibliography{psrrefs,rimrefs}

%% --------------------------------------------------------
\clearpage
%% --------------------------------------------------------
\clearpage
%% --------------------------------------------------------
\clearpage
%% --------------------------------------------------------
\clearpage
%% --------------------------------------------------------
\clearpage
%% --------------------------------------------------------
\clearpage
%% --------------------------------------------------------
\clearpage
%% --------------------------------------------------------
\clearpage
%% --------------------------------------------------------
\clearpage
%% --------------------------------------------------------
\clearpage
%% --------------------------------------------------------
\clearpage
%% --------------------------------------------------------
\clearpage
%% --------------------------------------------------------
\clearpage
\begin{deluxetable}{ccccccc}
\tabletypesize{\footnotesize}
\tablewidth{469.0pt}
\tablecaption
{
Summary of  {\em{RXTE}} Observations
\label{table1}
}
\tablehead
{
\colhead{Observing} &
\colhead{Typical} &
\colhead{Typical} &
\colhead{Number of} &
\colhead{Total} \\
\colhead{Cycle} &
\colhead{Exposure\tablenotemark{a}} &
\colhead{Separation\tablenotemark{a}} &
\colhead{Obs.\tablenotemark{b}} &
\colhead{Exposure\tablenotemark{c}} &
\colhead{First MJD $-$ Last MJD\tablenotemark{d}} &
\colhead{First Date $-$ Last Date} \\
&
\colhead{(ks)} &
\colhead{(weeks)} &
&
\colhead{(ks)} &
}
\startdata
1 & 11 & 0.1 & 4 & 45 & 50170.469$-$50171.904 & 29/03/1996$-$29/03/1996\\
2 & 1 & 4 & 14 & 16 & 50411.684$-$50795.523 & 24/11/1996$-$13/12/1997\\
3 & 20 & {\ldots} & 1 & 20 & 50893.083$-$50893.083 & 21/03/1998$-$21/03/1998\\
4 & {\ldots} & {\ldots} & 0 & 0 & {\ldots} & {\ldots} \\
5 & 3 & 4 & 15 & 46 & 51610.617$-$51950.256 & 07/03/2000$-$10/02/2001\\
6 & 7 & 6 & 10 & 54 & 51986.347$-$52282.163 & 18/03/2001$-$08/01/2002\\
7 & 15 & 6 & 9 & 100 & 52339.621$-$52634.456 & 06/03/2002$-$26/12/2002\\
8 & 5 & 2 & 21 & 124 & 52726.197$-$53046.235 & 28/03/2003$-$11/02/2004\\
9 & 5 & 2 & 27 & 86 & 53066.586$-$53420.507 & 02/03/2004$-$19/02/2005\\
10 & 5 & 2 & 36 & 120 & 53438.151$-$53787.328 & 09/03/2005$-$21/02/2006\\
\enddata
\tablenotetext{a}
{
The exposure and separation are only approximate. 
}
\tablenotetext{b}
{
When the last digit of the observation ID of two successive data sets
is different, the two data sets are considered separate observations.
}
\tablenotetext{c}
{
The total exposure does not include Earth occultation periods. 
}
\tablenotetext{d}
{
First MJD and Last MJD are the epochs, in Modified Julian Days, 
of the first and the last observations in a Cycle.
}
\end{deluxetable}
%% --------------------------------------------------------
\clearpage
\begin{deluxetable}{lccc}
\tabletypesize{\small}
\tablewidth{385.0pt}
\tablecaption
{
Spin Parameters for 4U~0142+61\tablenotemark{a}
\label{table2}
}
\tablehead
{
&
\colhead{Pre-Gap Ephemeris\tablenotemark{b}} &
\colhead{Post-Gap Ephemeris} &
\colhead{Possible Ephemeris\tablenotemark{c}} \\
\colhead{Parameter} &
\colhead{Spanning} &
\colhead{Spanning} &
\colhead{Spanning} \\
&
\colhead{Cycles 1 to 3} &
\colhead{Cycles 5 to 10} &
\colhead{All Cycles}
}
\startdata
MJD range & 50170.693$-$50893.288 & 51610.636$-$53787.372 & 50170.693$-$53787.372 \\
TOAs & 19 & 118 & 137 \\
$\nu$ (Hz) & 0.115099566(3) & 0.1150969337(3) &0.1150969304(2) \\
$\dot{\nu}$ (10$^{-14}$ Hz s$^{-1}$) & $-$2.659(3) & $-$2.6935(9) & $-$2.6514(7) \\
$\ddot{\nu}$ (10$^{-23}$ Hz s$^{-2}$) & --- & 0.417(10) & $-$1.7(2) \\
$d^{3}\nu/dt^{3}$ (10$^{-31}$ Hz s$^{-3}$) & --- & --- & 3.62(12) \\
$d^{4}\nu/dt^{4}$ (10$^{-39}$ Hz s$^{-4}$) & --- & --- & 8.7(3) \\
$d^{5}\nu/dt^{5}$ (10$^{-46}$ Hz s$^{-5}$) & --- & --- & $-$5.01(13) \\
$d^{6}\nu/dt^{6}$ (10$^{-54}$ Hz s$^{-6}$) & --- & --- & 6.6(4) \\
Epoch (MJD) & 50530.000000 & 51704.000025 & 51704.000000 \\
RMS residual & 0.019 & 0.023 & 0.019 \\
\enddata
\tablenotetext{a} { Numbers in parentheses are TEMPO-reported 1$\sigma$
uncertainties. } 
\tablenotetext{b} { The pre-gap ephemeris reported here is
slightly different from that reported in \citet{0142previous} because
here we take into account Cycle~1 and~3 observations. 
Note that both ephemerides return the same number of pulsar rotation 
cycles between the first and last pre-gap observations used by
\citet{0142previous}. }
\tablenotetext{c} { It is possible to find a different overall ephemeris
after adding an arbitrary but constant time jump to all post-gap TOAs. }
\end{deluxetable}
%% --------------------------------------------------------
\clearpage
%% --------------------------------------------------------
\end{document}